# RADIATIVE CAPTURE IN $^4$He$^{12}$C CHANNEL OF $^{16}$O NUCLEUS IN THE POTENTIAL CLUSTER MODEL

## S.B. Dubovichenko

Radiative capture processes in $^4$He$^{12}$C channel of $^{16}$O nucleus are considered on the basis of potential two-cluster model with intercluster interactions which contain forbidden states and reproduce phase shifts of elastic scattering and some characteristics of bound states. Astrophysical S-factors are calculated at low energies. It was shown that the model used enables to describe the total cross-sections of the photoprocesses.

Earlier in [1] the total cross-sections of the direct cluster photodisintegration and radiative capture processes in $^4$He$^3$H, $^4$He$^3$He, d$^4$He and $^3$He$^3$H channels of $^6$Li, $^7$Li and$^7$Be light cluster nuclei were studied on the basis of a potential two-cluster model with intercluster interactions which contain forbidden states (FS) [2]. Such interactions describe experimental phase shifts of elastic scattering at low energies (up to 20 - 40 MeV) and the FSs allow to take into account the Pauli exclusion principle in cluster interactions [3] without an attractive core. It was shown in [2] that they enable to reproduce some characteristics of bound states (BS) of $^6$Li and $^7$Li in cluster models where the channel clustering probability is comparatively high. Orbital states in such systems turn out to be pure according to Young schemes, so potentials obtained on the basis of experimental data on elastic scattering phase shifts can be used for calculation of the nucleus ground state characteristics [2].

Calculations of the total cross-sections of photoprocesses on lighter systems of Nd, p$^3$H, n$^3$He, dd and d$^3$He type were made in [4] using a potential cluster model with forbidden states. Situation in these systems is more complicated because of orbital symmetry mixing in the minimum spin channels. Therefore, it is necessary to extract a pure component from interactions obtained on the basis of experimental elastic scattering phase shifts which can be used for analysis of the ground state characteristics [5, 6]. Orbital schemes mixing in the minimum spin states exists not only for the most of the lightest cluster systems but for some heavy N$^6$Li, N$^7$Li and d$^6$Li systems as well [7].

Differential cross-section calculations of photoprocesses in p$^3$H and dd channels for potentials with FSs and orbital schemes separation were made successfully in [6]. Differential cross-sections for $^6$Li and $^7$Li cluster channels in two-cluster models with FSs were considered in [8]. The total cross-sections were analyzed on the basis of resonating group method (RGM) [9]. Calculations of the total cross-sections of photoprocesses in $^6$Li nucleus in three-body $^4$Henp model with FSs were made in [10]. Total cross-sections of $^4$Hed capture in two-cluster model were calculated [11].

The photo-processes in $^4$He$^{12}$C channel of $^{16}$O were studied in [12] in terms of the two-cluster model with deep intercluster interaction. However, forbidden state structure was not investigated for potentials corresponding to different partial waves. So, it appears to be interesting to reconsider the process and to analyze again all possible EJ transitions to different $^{16}$O levels in the framework of the cluster model for potentials with FSs and to analyze FS location.

Let us consider now orbital states classification in $^4$He$^{12}$C system with spin S and isospin T equal to zero. The possible orbital Young schemes in N=n$_1$+n$_2$ particles system can be obtained according to the Littlewood theorem [13] as a direct external product of orbital schemes for each sub-system. In this case we have: {f} = {444} $\otimes$ {4} = {844} + {754} + {7441} + {664} + {655} + {6541} + {6442}+ {5551} + {5542} + {5443} + {4444}. Here

{4} and {444} schemes correspond to $^4$He and $^{12}$C nuclei in the ground state (GS). There is only one orbital scheme {4444} which can be allowed in this case in accordance to known rules [13]. The other orbital configurations are forbidden. Particularly, all possible configurations where the number of cells in Young schemes is more than four can not be realized. So, there could be more than four nucleons in the S-shell of the nucleus. It is possible to specify orbital moments for different Young schemes using Elliot Rule [13]. L=0 orbital moment realized for the following orbital schemes: {4444}, {5551}, {664}, {844}, {6442}. So, there are four FSs and one allowed state (AS) with {4444} in the bound S-state. This result can be used for the qualitative estimation of the number of bound states. As the potential with three BSs was considered in [12], and we will consider only the interactions with four and five BSs. The spectrum of $^{16}$O nucleus levels is shown in Fig.1a.

To calculate the capture cross-section in the long-wave approximation, we used a well-known expression [1,4,9]

$$\sigma_c(NJ) = \frac{8\pi \, K^{2J+1}}{\hbar^2 q^3} \frac{\mu}{(2S_1+1)(2S_2+1)} \frac{J+1}{J[(2J+1)!!]^2} \sum_{\substack{L_i, L_f \\ J_i, J_f}} |T_J(N)|^2,$$

(1)

where N = E or M and

$$T_J(E) = A_J \, I_J \, P_J + (\, B_{1J} \, N_{1J} + B_{2J} \, N_{2J} \,) \, I_J = T_J(EL) + T_J(ES)$$
$$T_J(M) = C_J \, I_{J-1} \, G_J + (\, D_{1J} \, N_{1J} + D_{2J} \, N_{2J} \,) \, I_{J-1} = T_J(L) + T_J(S)$$

Electromagnetic operators giving $T_J$ values in the cluster model are given in the form [1]

$$Q_{Jm}(L) = e \, \mu^J \left[ \frac{Z_1}{M_1^J} + (-1)^J \frac{Z_2}{M_2^J} \right] R^J Y_{Jm} = A_J R^J Y_{Jm},$$

$$Q_{Jm}(S) = -\frac{e\hbar}{m_0 c} K \left[ \frac{J}{J+1} \right]^{1/2} \left[ \mu_1 \hat{S}_1 \frac{M_2^J}{M^J} + (-1)^J \mu_2 \hat{S}_2 \frac{M_1^J}{M^J} \right] R^J Y_{Jm}^J =$$

$$= (B_{1J} \hat{S}_1 + B_{2J} \hat{S}_2) R^J Y_{Jm}^J,$$

(2)

$$W_{Jm}(L) = i \frac{e\hbar}{m_0 c} \frac{\sqrt{J(2J+1)}}{J+1} \left[ \frac{Z_1}{M_1} \frac{M_2^J}{M^J} + (-1)^{J-1} \frac{Z_2}{M_2} \frac{M_1^J}{M^J} \right] R^{J-1} \hat{L} Y_{Jm}^{J-1} =$$
$$= C_J R^{J-1} \hat{L} Y_{Jm}^{J-1},$$

$$W_{Jm}(S) = i \frac{e\hbar}{m_0 c} \sqrt{J(2J+1)} \left[ \mu_1 \hat{S}_1 \frac{M_2^{J-1}}{M^{J-1}} + (-1)^{J-1} \mu_2 \hat{S}_2 \frac{M_1^{J-1}}{M^{J-1}} \right] R^{J-1} Y_{Jm}^{J-1} =$$

$$= (D_{1J} \hat{S}_1 + D_{2J} \hat{S}_2) R^{J-1} Y_{Jm}^{J-1}.$$

Here, J is the multipole order; q is the wave number of the cluster relative motion; $m_0$ is the nucleon mass; $\mu$ is the nucleus reduced mass; $M_i$, $Z_i$, $S_i$ and L are masses, charges, spins, and orbital angular momenta of the clusters; $\mu_i$ are magnetic moments of the clusters; K is the wave number of the photon; M is the nuclear mass; and R is the intercluster distance.

The general formulae for $P_J$, $N_J$ and $G_J$ are given in [1]. The $I_J$ are integrals having the form

$$I_J = < L_f | R^J | L_i >,$$

where $| L_{i,f}>$ are the radial WF of the initial and final states. The Gauss potentials of the intercluster interaction have the form [4]

$$V(r) = - V_0 \exp(- r^2 / R_0^2 ) + V_c \qquad (3)$$

where $V_c$ is the Coulomb interaction for a charged sphere with a radius of $R_c$=3.55 Fm [12].
It is impossible in the considered system to obtain potentials enabling to describe both phase shifts of scattering and GS characteristics, as it was in the case of lighter cluster configurations. Therefore, it appears that there are two groups of interactions not related to each other. The ground state potential was made up according to the requirements of description of such characteristics as the bound energy, charged radius, Coulomb formfactor at a low momentum transferred and that the number of BSs should satisfy the above-mentioned classification. Such a potential should contain three or four FSs and one AS at the binding energy of $^{16}$O nucleus in $^4$He$^{12}$C channel. Interactions 1 and 2 from Tab. 1 satisfy these conditions giving the charge radius at 2.66 Fm and at 2.72 Fm while the experimental value is 2.710 ≀ 0.015 Fm [14]. These interactions describe well enough the Coulomb formfactor up to the transferred momentum of 1.5 Fm$^{-1}$, as one can see from Fig.1b (the dotted curve is for the interaction 1 and the solid line is for the interaction 2). Experimental [14] and calculated energy levels and the FS energies are given in Tab. 1. The potential from [12] ($V_0$=110 MeV and $R_0$=2.3 Fm) which contains two FSs leads to the radius of 2.6 Fm and the formfactor lays a bit above the experimental data, as it is can be seen from Fig. 1b (dashes curve). The experimental formfactor of $^{16}$O is taken from [15].

The method described in [2] was used for the formfactor calculations and the following expression was applied to parameterization of formfactors of $^4$He and $^{16}$O clusters:

$$F = ( 1 - ( aq^2)^n ) \exp(-bq^2) \qquad (4)$$

where $a$=0.09986 Fm$^2$, $b$=0.46376 Fm$^2$, $n$=6 [2] - for $^4$He and $a$=0.31 Fm$^2$, $b$=1.18 Fm$^2$, $n$=4.5 - for $^{12}$C. Experimental data from [16].

The potentials of other bound states which spectrum of is shown in Fig. 1a are made up according to the requirement of description of reduced probabilities of electromagnetic transitions between different levels. The results of the reduced probability calculations are shown in Tab. 2 and the parameters of such interactions and the BS energies are given in Tab. 1 (no. 3-6). The experimental data are from [14, 17].

Table 1. The potentials of interactions for $^4$He$^{12}$C system. $E_{th}$ are the calculated and $E_{exp}$ are the experimental energy levels. $E_{bs}$ are the energies of bound forbidden states.

| N | L | $V_0$ (MeV) | $R_0$ (Fm) | $E_{th}$ (MeV) | $E_{exp}$(MeV) | $E_{bs}$ (MeV) |
|---|---|---|---|---|---|---|
| 1 | 1S | 176.8 | 2.3 | 7.17 | 7.162 | 35.9; 76.5; 126.6 |

| 2 | 1S | 256.65 | 2.3 | 7.16 | | 37.5; 80.7; 134.3; 197.0 |
|---|----|--------|-----|------|---|--------------------------|
| 3 | 2S | 97.78 | 3.0 | 1.11 | 1.113 | 16.0; 38.3; 66.2 |
| 4 | 1P | 104.1 | 2.5 | 0.048 | 0.045 | 19.2; 48.6 |
| 5 | 1D | 88.53 | 3.2 | 0.246 | 0.245 | 13.7; 33.5 |
| 6 | 1F | 191.42 | 1.9 | 1.03 | 1.032 | 38.3 |
| 7 | S | 90.0 | 2.3 | | | 20.7; 52.9 |
| 8 | S | 155.0 | 2.3 | | | 1.37; 25.2; 61.5; 107.7 |
| 9 | P | 145.0 | 2.5 | | | 13.6; 42.2; 79.7 |
| 10 | D | 254.8 | 1.3 | | | 57.0 |
| 11 | D | 434.9 | 1.3 | | | 61.8; 166.9 |
| 12 | F | 140.0 | 2.6 | | | 11.8; 39.4 |
| 13 | G | 111.15 | 2.8 | | | 13.6 |

The bound states of the $^4$He$^{12}$C system are marked by figures which show the level numbers and by letters which correspond to the orbital moments (see Tab. 1 and Fig.1a). Tab. 2 shows that potentials used enable to describe experimental data on reduced probabilities of the electromagnetic transitions. The 1D and 2S interactions do not differ much from potentials from [12] and this leads to similar results for the 1D } 2S transition probability. The value of 63 e$^2$ Fm$^4$ was obtained for this characteristic in the work [12].

Fig. 1a. - spectrum of $^{16}$O nucleus.

Table 2. The calculated and experimental reduced probabilities of electromagnetic transitions in $^4$He$^{12}$C system.

| $L_i \longrightarrow L_f$ | $B_{th}$ (e$^2$ Fm$^4$) | $B_{th}$ (e$^2$ Fm$^4$) | $B_{exp}$ (e$^2$ Fm$^4$) |
|---------------------------|-------------------------|-------------------------|--------------------------|

|          | (Potential 1) | (Potential 2) | [14,17]            |
|----------|---------------|---------------|--------------------|
| 1F—> 1S  | 15.0          | 11.8          | 13.9(1.2); 14.6(1.5) |
| 1F—> 2S  | 7.0           | 7.0           | --                 |
| 1D—> 1S  | 6.6           | 14.8          | 4.6 - 7.9          |
| 1D—> 2S  | 67.1          | 67.1          | 63.0; 71(8)        |

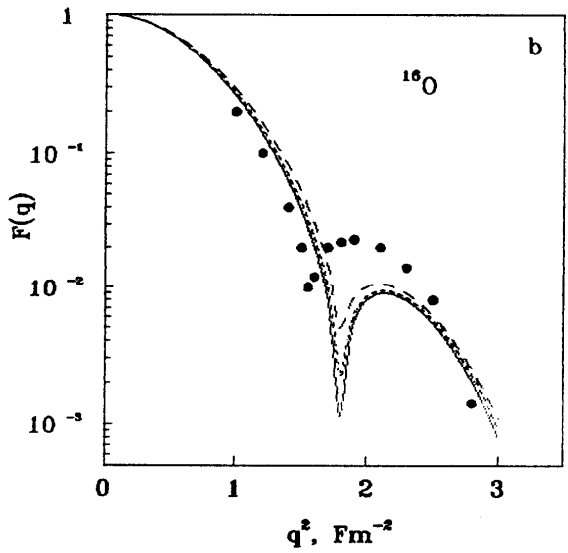

Fig. 1b - the Coulomb formfactor of $^{16}$O nucleus, experimental data from [15]. The solid line is the calculation results for potential 2 from table 1; dotted curve is for potential 1, dashes curve is for the potential from [12].

The potentials describing the experimental scattering phase shifts and differing from interactions for BSs are given in Tab. 2 under the numbers from 7 to 13. Fig. 2 and 3 shows calculation results for scattering phase shifts together with experimental data [18]. There are two types of interactions for S and D waves which give the same results but contain different numbers of FSs. The potentials for P and G waves are practically equal to the potentials given in [12].

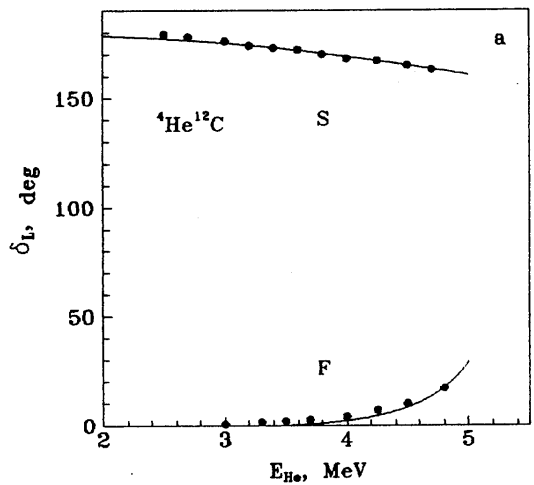

Fig. 2. Phase shifts of the elastic $^{4}$He$^{12}$C scattering: a - S and F phase shifts, b - P phase shift. Experimental data from [18].

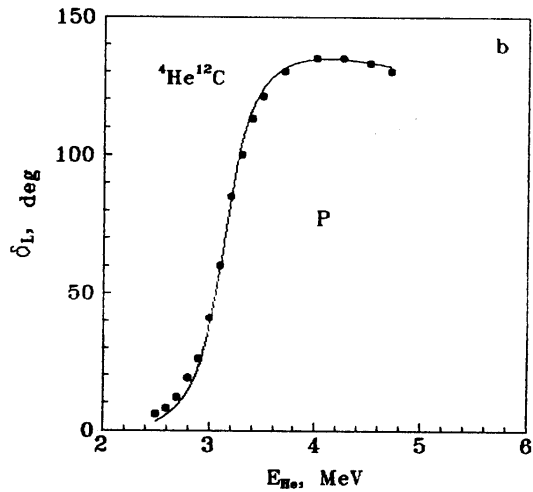

Considering the radiative capture processes, the possibility of E2 and E3 transitions is taken into account. Dipole transitions in such a model are forbidden because of $(Z_1/M_1 - Z_2/M_2)$ factor equals to zero. The capture cross-section experimental data for transitions to the ground 1S state and to the binding 1D level of $^{16}O$ nucleus were obtained in work [19].

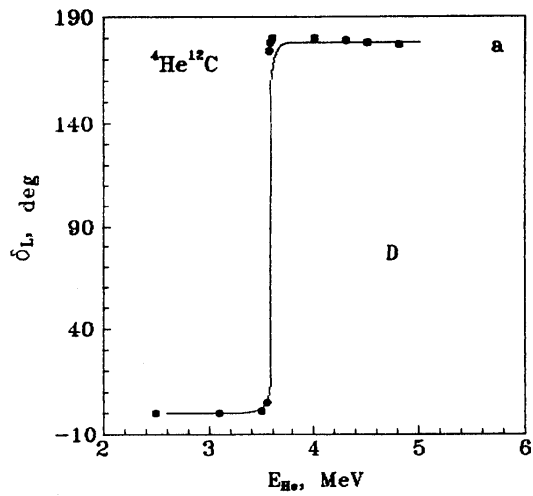

Fig. 3. Phase shifts of the elastic $^4He^{12}C$ scattering: a - D and b - P phase shifts. Experimental data from [18].

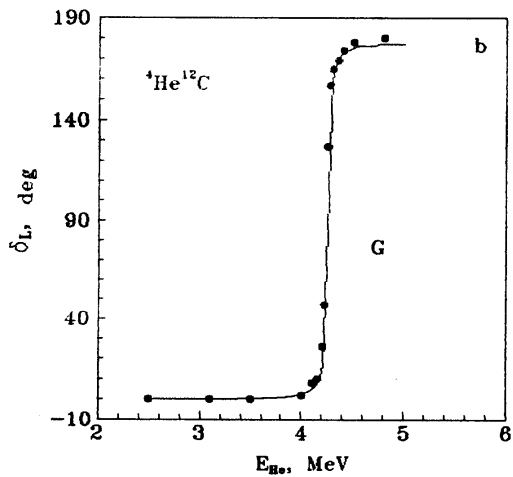

Experimental cross-sections of transitions to the ground states have a clear maximum at an energy about 2.4 MeV which corresponds to $1^+$ resonance at 2.46 MeV. Such shape of the cross-section indicates presence of E1 process which is not contained in the considered model. However, there is a peak at 2.69 MeV in the cross-section in Fig. 4a corresponding to $2^+$ resonance (see Fig. 1a). The peak is the result of a possible E2 transition D --> 1S. Results of the process cross-section calculation with the above potentials and with the first variant of GS interaction are shown in Fig. 4a by a solid line. The dashed line shows the calculated cross-section for E3 transition from F scattering wave to the ground state. Calculation results for the GS potential 2 are shown by a dotted curve. This cross-section is practically explain the experimental data at energies lower than 2 MeV and the position of $2^+$ peak. To obtain a more complete picture of the process it is necessary to consider also E1 cross-section as it can change the total cross-section. Fig. 4b shows the astrophysical S factor for E2 transition to the GS. Its value is tending to 0.01 MeV at 300 keV for GS potential 1 (solid line) and to 0.02 MeV for variant 2 (dashes curve). This is, at least, one order lower than the data which presented in [12]. Evidently, only E1 transition P ↾ 1S has the main role in the cross-section and S-factor calculations.

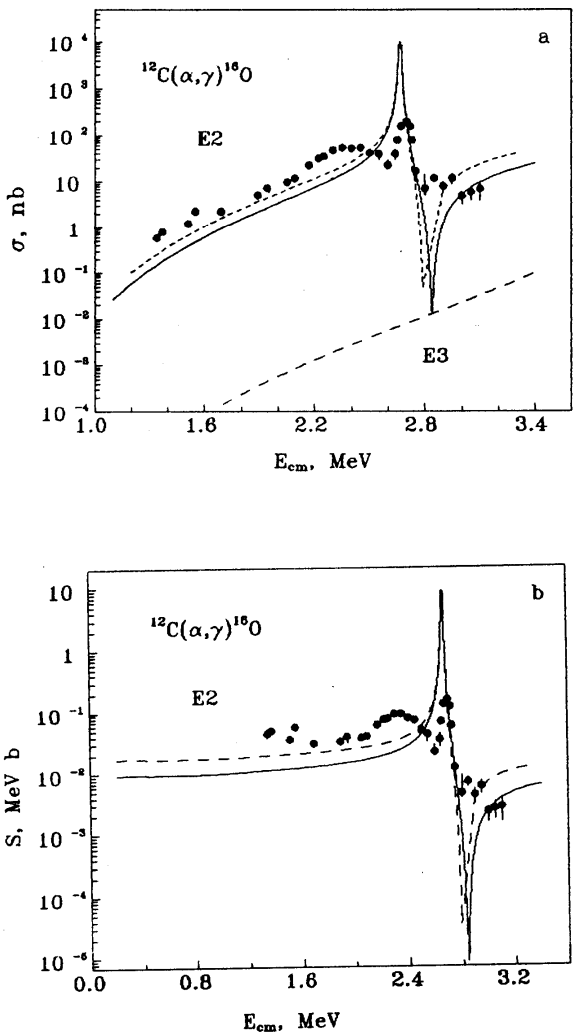

Fig. 4. a - the total cross-sections of $^4$He$^{12}$C radiative capture to the ground state of $^{16}$O nucleus, b - astrophysical S-factors. Experimental data from [19]. The lines are the results of calculations with potentials from Table 1.

There are several E2 processes (such as 1. - G --> 1D;  2. - D --> 1D;  3 - S --> 1D) which are the result of transitions to the 1D state. Fig. 5a shows the calculated cross sections corresponding to these processes (1 - double-dot-and-dash line, 2 - short-dashed line, 3 - long-dashed line). The total cross-section is shown by dotted curve. It is clear that the considered transitions practically describe the experimental data on position and height of peaks at 2.69 MeV and 3.19 MeV corresponding to the D and G resonances. However, there is a small peak in the experimental cross-sections at 2.46 MeV which obviously is the result of the transition from P scattering wave. At the same time, if we assume that the experimental cross-section includes the transition to 1P level at -0.045 MeV (this value is differ from 1D state only at 0.2 MeV) it means that it is possible to investigate E2 process of P ↾ 1P type. The cross-section of such a process is shown in Fig. 5a by the dot-and-dash curve. The total cross-section which takes into account this transition is shown by a solid line and this curve reproduces practically the form of the experimental cross-sections. S-factor of this process is shown in Fig. 5b by a solid line. It is equal to 0.001 MeV b at an energy of 300 keV and then it falls down at lower energies.

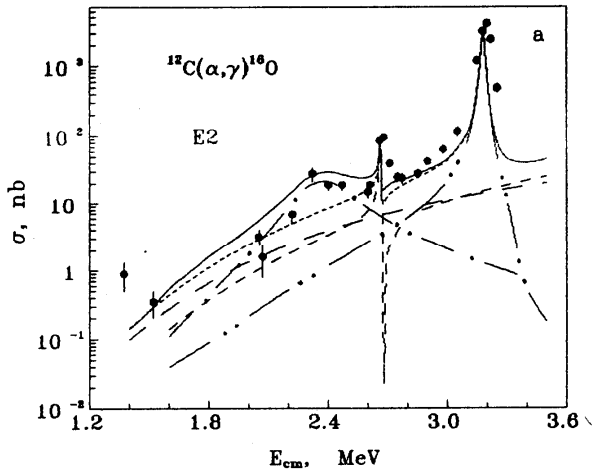

Fig. 5. a - the total cross-sections of $^4He^{12}C$ radiative capture to the $2^+$ resonance level of $^{16}O$ nucleus, b - astrophysical S-factors. Experimental data from [19]. The lines are the results of calculations of different transitions with potentials from Table 1.

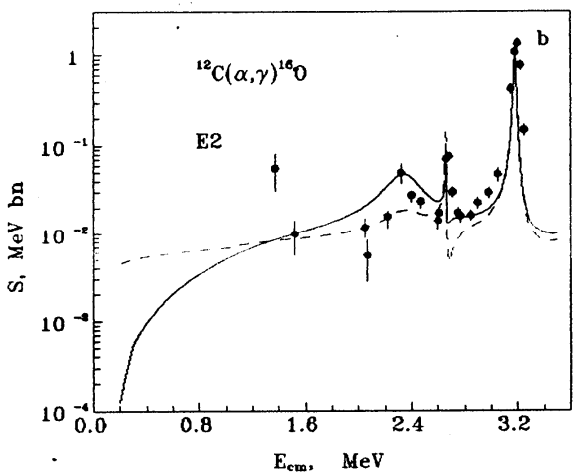

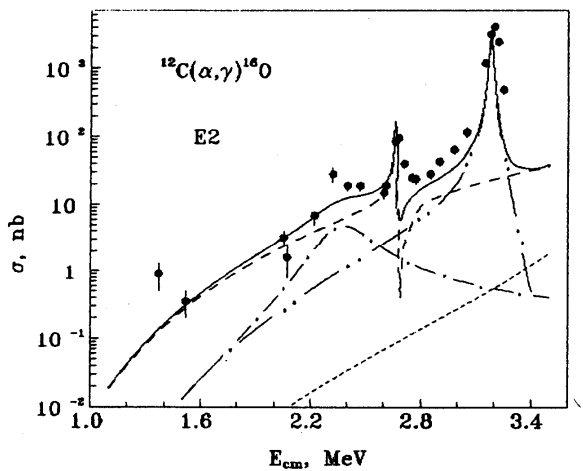

It is interesting to note that if we consider such transitions as 4.- F--> 1F; 5.- D--> 2S and 6.- P--> 1F instead of D--> 1D and S--> 1D transitions, it will be possible to describe the general form of the experimental cross-sections (see Fig. 6). The point line is transition no. 4, the dashed line is transition no. 5, dot-and-dashed line is transition no. 6 and double-dot-and-dash line is transition no.1. Solid line is total cross section. The S-->1D process makes the main contribution to cross-section at low energies in Fig. 5a, but in this case the D-->2S transition makes the main contribution and the role of other transitions is not important. The corresponding S-factor is shown in Fig. 5b by the dashed line. It lays slightly higher than the S-factor for 1D level and it is equal to 0.005 MeV b at the energy of 300 keV.

Fig. 6. The total cross-sections of $^4He^{12}C$ radiative capture to different levels of $^{16}O$ nucleus. Experimental data from [19]. The lines are cross-sections of different transitions with potentials from Table 1.

It is clear from the above-mentioned results that the used potentials of $^4He^{12}C$ channel of $^{16}O$ nucleus which are in agreement with the elastic scattering phase shifts and with the probabilities of radiative transitions for BSs allow to describe the radiative cross-section of capture to $2^+$ level on the base of E2 transitions. Capture cross-sections to the GS, in general, are described well enough only at the range of $2^+$ resonance and below 2 MeV. The main contribution to the cross-section at other energies is made by the E1 (P--> 1S) transition which is absent in this cluster model. It is difficult to give an only single conclusion about the form of GS potential as both of the considered interactions lead to the similar results.